\documentclass[sigconf]{acmart}

\renewcommand\footnotetextcopyrightpermission[1]{} 
\usepackage{color}
\usepackage{url}
\usepackage{verbatim} 
\usepackage{subfig} 
\usepackage{multirow}
\usepackage{algorithm}
\usepackage[noend]{algorithmic}
\usepackage{algorithmic}
\usepackage{paralist}
\usepackage{xspace}
\usepackage{graphicx}
\usepackage{epsfig}
\usepackage{amsmath}
\usepackage{amssymb}

\apptocmd{\thebibliography}{\normalsize}{}{}

\newcommand{\hide}[1]{}
\newcommand{\xhdr}[1]{\vspace{1.7mm}\noindent{{\bf #1.}}}

\newcommand{\board}{{\ensuremath{b}}}
\newcommand{\Boards}{{\ensuremath{B}}}
\newcommand{\pin}{{\ensuremath{p}}}
\newcommand{\Pins}{{\ensuremath{P}}}

\newcommand{\eg}{\emph{e.g.}}
\newcommand{\ie}{\emph{i.e.}}

\graphicspath{{./FIG/}}

\acmDOI{10.475/123_4}

\acmISBN{123-4567-24-567/08/06}

\acmYear{2018}
\copyrightyear{2018}

\begin{document}

\title{Pixie: A System for Recommending 3+ Billion Items to\\ 200+ Million Users in Real-Time}

\author{Chantat Eksombatchai, Pranav Jindal, Jerry Zitao Liu, Yuchen Liu,\\ Rahul Sharma, Charles Sugnet, Mark Ulrich, Jure Leskovec 
}
\affiliation{
	\institution{Pinterest}
}
\email{{pong, pranavjindal, zitaoliu, yuchen, rsharma, sugnet, mu, jure}@pinterest.com}


\begin{abstract}
User experience in modern content discovery applications critically depends on high-quality personalized recommendations. However, building systems that provide such recommendations presents a major challenge due to a massive pool of items, a large number of users, and requirements for recommendations to be responsive to user actions and generated on demand in real-time. Here we present Pixie, a scalable graph-based real-time recommender system that we developed and deployed at Pinterest. Given a set of user-specific pins as a query, Pixie selects in real-time from billions of possible pins those that are most related to the query. To generate recommendations, we develop Pixie Random Walk algorithm that utilizes the Pinterest object graph of 3 billion nodes and 17 billion edges. Experiments show that recommendations provided by Pixie lead up to 50\% higher user engagement when compared to the previous Hadoop-based production system. Furthermore, we develop a graph pruning strategy at that leads to an additional 58\% improvement in recommendations. Last, we discuss system aspects of Pixie, where a single server executes 1,200 recommendation requests per second with 60 millisecond latency. Today, systems backed by Pixie contribute to more than 80\% of all user engagement on Pinterest.
\end{abstract}

\maketitle

%
%

\section{Introduction}
\label{sec:intro}

Pinterest is a visual catalog with several billion {\em pins}, which are visual bookmarks containing a description, a link, and an image or a video.
A major problem faced at Pinterest is to provide personalized, engaging, and timely recommendations from a pool of 3+ billion items to 200+ million monthly active users.

Recommendations at Pinterest are a problem with a scale beyond the classical recommendation problems studied in the literature. Pinterest has a catalog of several billion pins that the recommender system can choose from. However, classical recommender systems that consider catalogs that only contain millions of items 
(movies~\cite{bennett2007netflix,koren2009matrix}, videos~\cite{baluja2008video,davidson2010youtube,youtube}, friends to follow~\cite{backstrom2011supervised,goel2015follow,wtf}). In contrast, Pinterest recommends from a catalog of billions of items, which makes the recommendations problem much more challenging.

A second important challenge is posed by the fact that recommendations have to be calculated on demand and in real-time. This real-time requirement (\ie, sub 100 millisecond latency per recommendation request) is crucial for two reasons: (1) 
Users prefer recommendations responsive to their behavior, thus recommendations have to be computed on demand and in real-time so the system can instantaneously react to changes in user behavior and intent;
(2) 
Real-time requirement also brings a drastic change in the design of the entire system. For example, even if recommendations would only take one second to compute, such times are too long for the user to wait. In turn, this would mean that recommendations for all users would have to be precomputed and materialized on a daily schedule. Moreover, the total number of registered users is usually much larger than the number of daily active users, so a lot of time and resources would be wasted updating recommendations for inactive users.

\xhdr{Present work: Pixie}
Here we present Pixie, a scalable real-time graph-based recommendation system deployed at Pinterest. Currently, pins recommended by Pixie represent more than 80\% of all user engagement 
at Pinterest. In an A/B tests recommendations provided by Pixie increase per pin engagement by up to  50\% higher compared to the previous Pinterest recommendation systems.

Users at Pinterest view pins and curate them into collections called {\em boards}. This way a single pin can be saved by thousands of users into tens of thousands of different boards. For example, the same recipe pin could be saved by different users to several different boards such as ``recipes'', ``quick to cook'', ``vegetarian'', or ``summer recipes.'' This manual curation mechanism provides a great source for recommendations, because curation captures the multi-faceted relationships between objects. With hundreds of millions of users manually categorizing/classifying pins into boards we obtain an object graph of multi-faceted relationships between pins. Thus, we can think of Pinterest as a giant human curated bipartite graph of 7 billion pins and boards, and over 100 billion edges~\footnote{There are over 100 billion pins in all boards at Pinterest, represented by the 100 billion edges in our graph. The unique number of pins is smaller because the same pin is usually saved to many boards.}.

We use this bipartite graph of pins and boards to generate recommendations. As a user interacts with pins at Pinterest our method uses these pins to create a query set $Q$ of pins each with its own weight. The query set is user-specific and changes dynamically---it can contain most recently interacted pins as well as pins from long time ago. Given the query $Q$, we then generate recommendations using Pixie Random Walk algorithm. Because the walk visits both boards as well as pins both types of objects can be recommended to the user. Furthermore, the algorithm is fast, scalable, and runs in constant time that is independent of the input graph size.

Our novel Pixie Random Walk algorithm includes the following innovations, which are critical for providing high-quality recommendations: 
(1) We bias the Pixie Random Walk in a user-specific way, for example, based on the topic and language of the user; 
(2) We allow for multiple query pins with different importance weights, which allows us to capture entire context of the user's previous behavior; 
(3) Our method combines results from multiple independent random walks in such a way that it rewards recommendations that are related to multiple query pins. In combinations with (2) this leads to more relevant recommendations; 
(4) Our Pixie Random Walk uses special convergence criteria which allows for early stopping and is crucial for achieving real-time performance and throughput;
Last, (5) our Pixie algorithm allows for recommending both pins as well as boards, which helps solving the cold-start problem. To recommend fresh new pins Pixie first recommends boards (rather than pins) and then serves the new pins saved to those boards.
In addition, we also develop a graph curation strategy that further increases the quality of recommendations by 58\%. This curation also lowers the size of the graph by a factor of six which further improves runtime performance of Pixie.

Our Pixie algorithm has several important advantages. The algorithm offers the flexibility to dynamically bias the walk. For example, in Pixie we bias the walk to prefer to recommend content local to the language of the user, which boosts user engagement. Pixie allows for computing recommendations based on multiple query pins. Furthermore, we can vary the walk length to make trade-offs between broader and narrower recommendations. For areas of Pinterest intended to provide unexpected, exploratory recommendations Pixie can walk farther in the graph for more diverse recommendations. On the other hand, to generate narrowly focused and topical recommendations, Pixie can use shorter walks. Pixie Random Walk also has several advantages over traditional random walks (or over simply counting the number of common neighbors): In classical random walks low degree nodes with fewer edges contribute less signal. This is undesirable because smaller boards (lower degree nodes) tend to be more topically focused and are more likely to produce highly relevant recommendations. In Pixie Random Walk we solve this by boosting the impact of smaller boards. And last, Pixie Random Walk is efficient and runs in constant time that is independent of the input graph size.

Deployment of Pixie is facilitated by the availability of large RAM machines. In particular, we use a cluster of Amazon AWS r3.8xlarge machines with 244GB RAM. We fit the pruned Pinterest graph of 3 billion nodes and 17 billion edges into about 120 GB of main memory. This gives several important benefits: (1) Random walk does not have to cross machines which brings huge performance benefits; (2) The system can answer queries in real-time and multiple walks can be executed on the graph in parallel; And, (3) Pixie can be scaled and parallelized by simply adding more machines to the cluster. 

Overall, Pixie takes less than 60 milliseconds (99-percentile latency) to produce recommendations. Today, a single Pixie server can serve about 1,200 recommendation requests per second, and the overall cluster is serving nearly 100,000 recommendation requests per second. 
Pixie is written in C++ and is built on top of the Stanford Network Analysis Platform (SNAP)~\cite{snap}.

The remainder of this paper is structured around the primary contributions of our work. After a brief discussion of the related work in Section~\ref{sec:related}, we explain
the Pixie Random Walk algorithm in Section~\ref{sec:pixie}. 
We discuss graph pruning in Section~\ref{sec:graph_curation} and the system implementation in Section~\ref{sec:impl}. Section~\ref{sec:experiments} evaluates the system as well as the recommendations. Section~\ref{sec:discussion} discusses Pixie's impact and use cases at Pinterest. Finally, we conclude in Section~\ref{sec:conclusion}.

\section{Related Work}
\label{sec:related}

Recommender systems are a large and well-investigated research field. Here we break the related work into several lines and focus in large-scale industrial recommeder systems. 

\xhdr{Web-scale recommender systems}
Several web-scale production systems have been described in the past~\cite{linkedin,youtube,amazon}. However, unlike Pixie, these systems are not real-time and their recommendations are precomputed. In practice, response times below 100 milliseconds are considered real-time because such systems can then be incorporated in the real-time serving pipeline. For example, if providing recommendations would take just 1 second, then the user would have to wait too long for the recommendations to be generated. In such cases recommendations would have to be precomputed (say once a day) and then served out of a key-value store. However, old recommendations are stale and not engaging. The real-time requirement is thus crucial because it allows the recommender to instantaneously react to changes in user behavior and intent. Responding to users in real-time allows for highly engaging and relevant recommendations. Our experiments show that reacting to user's intent in real-time leads to 30-50\% higher engagement than needing to wait days or hours for recommendations to refresh. 

Other examples of real-time recommendation systems include news recommendations~\cite{google,graphjet}. However, such systems recommend only the latest content. The major difference here is in scale---Pinterest's catalog contains 1,000-times more items than traditional recommender systems can handle.

\xhdr{Random-walk-based approaches}
Many algorithms  random walks to harness the graph structure for recommendations~\cite{backstrom2011supervised,Tong:2006:FRW:1193207.1193363,bahmani2010fast}. 
Perhaps, the closest to our work here is the ``who to follow'' systems at Twitter~\cite{wtf,goel2015follow} that places the entire follow graph in memory of a single machine and runs a personalized SALSA algorithm~\cite{salsa}. 
These types of Monte Carlo approaches measure the importance of one node relative to another, and recommendations can be made according to these scores~\cite{fogaras2005towards}. In contrast, we develop a novel random walk that is faster and provides better performance. 

\xhdr{Traditional collaborative filtering approaches}
More generally, our approach here is related to Collaborative filtering (CF) which makes recommendations by exploiting the interaction graph between users and items by matching users that have similar item preferences. CF relies on factorizing user-item interaction matrices to generate latent factors representing users and items \cite{konstan1997grouplens,sarwar2001item,zhuang2013fast,kabiljo_recommending_2015}.
However, the time and space complexity of factorization-based CF algorithms scale (at least) linearly with the number of nodes in the input user-item graph, making it challenging to apply these algorithms to problems containing billions of items and hundreds of millions of users. In contrast, our random walk based recommendation algorithm runs in constant time which is independent of the graph/dataset size.

\xhdr{Content-based methods}
In purely content-based recommender systems, representations for items are computed solely based on their content features~\cite{pazzani2007content}. Many state-of-the-art web-scale recommendation systems are content-based, often using deep neural networks~\cite{covington2016deep,van2013deep,cheng2016wide,zheng2017joint}. While these algorithms can scale to large datasets because the dimension of parameter space only depends on the dimension of feature space, these approaches have not leveraged information from the graph structure which (as our experiments show) is essential for Pinterest.

\section{Proposed Method}
\label{sec:proposed}

Pinterest is a platform where users interact with {\it pins}.
Users can {\it save}
relevant pins to {\it boards} of their choice. These boards are collections
of similar pins. For example, a user can create a board of recipes
and collect pins related to food items in it. Pinterest can be seen as a collection of boards
created by its users, where each board is a set of curated pins and each pin can be saved to hundreds of thousands of different boards.
More formally, Pinterest can be organized as an undirected bipartite graph $G=(\Pins,\Boards, E)$\footnote{We note there are alternative ways to define graph $G$. For example, one could also connect users to boards they own. Here we present this simplified graph but all our algorithms generalize to more complex graphs as well.}. Here, $\Pins$ denotes a set of pins and $\Boards$ denotes the set of boards.
The set $\Pins\cup\Boards$ is the set of nodes of $G$.
There is an edge $e\in E$ between a pin $\pin\in\Pins$ and
a board $\board\in\Boards$ if a user saved $\pin$ to $\board$.
We use $E(\pin)$ to denote board nodes connected to pin $\pin$ (and $E(\board)$ for pins connected to $\board$). We assume that $G$ is connected, which is also the case in practice.

On the input Pixie receives a weighted set of query pins $Q=\{(q, w_q)\}$, where $q$ is the query pin and $w_q$ is its importance in the query set. The query set $Q$ is user-specific and is generated dynamically after every action of the user---most recently interacted pins have high weights while pins from long time ago have low weights. Given the query $Q$ Pixie then generates recommendations by simulating a novel version of a biased random walk with restarts. 

\subsection{Pixie Random Walk}
\label{sec:pixie}

To ease the explanation of Pixie we first explain the basic random walk and then discuss how to extend it into a novel random walk algorithm used by Pixie. All the innovations on top of the basic random walk are essential for Pixie to achieve its full performance.

\hide {In this section, we assume that $G$ is connected, static, and fits in the memory of a single machine.
 ($G$ is updated once daily, Section~\ref{sec:graphupdate}.)
This graph is the main data structure that our recommendation system Pixie works with.
Today at Pinterest, there is a farm of many Pixie servers with each one holding the entire graph in memory and supporting up to 1,000 queries per second, with each query taking in a set of 1 - 50 pins as input and returning up to 2,000 pins as recommendations.
Pixie serves over 99\% of basic queries in less than 60 milliseconds.
This short latency is necessary to ensure that the recommendations are
delivered in real time. Section~\ref{sec:basic} below discusses
the basic random walk algorithm while Section~\ref{sec:earlystop} describes extensions to the basic algorithm and Section~\ref{sec:graph_curation} explains graph pruning.
} 

\xhdr{Basic Random Walk}
\label{sec:basic}
Consider the simple case when user-specific query $Q$ contains a single pin $q$.
Given, an input query pin $q$, one can simulate many short random walks on $G$, each starting from $q$, and record the {\em visit count} for each candidate pin $\pin$, which counts the number of times the random walk visited pin $\pin$. The more often the pin is visited, the more related it is to the query pin $q$.

The basic random walk procedure {\sc BasicRandomWalk} is described in Algorithm~\ref{alg:rw}~\cite{Tong:2006:FRW:1193207.1193363}.
Each random walk produces a sequence of {\it steps}. Each step is composed of three operations. First, given the current pin $\pin$ (initialized to $q$) we select an edge $e$ from $E(\pin)$ that connects $q$ with a board $\board$. Then, we select pin $\pin'$ by sampling an edge $e'$ from $E(\board)$ that connects $\board$ and $\pin'$. And third, the current pin is updated to $\pin'$ and the step repeats.

Walk lengths are determined by parameter $\alpha$. The total of the number of steps across all such short random walks determines the time complexity of this procedure and we denote this sum by $N$. Finally, we maintain a {\it counter} $V$ that maps candidate pins to the visit count. To obtain the recommended pins, we can extract the top visited pins from the returned counter and return them as the query response. The time taken by this procedure is constant and independent of graph size (determined by parameter $N$).

\begin{algorithm}[t]
\vspace{-0pt}{\sc BasicRandomWalk}($q$: Query pin,
			      $E$: Set of edges,
			      $\alpha$: Real,
                  $N$: Int) \\
 \begin{algorithmic}[1]
    \STATE totSteps = 0, $V=\vec{0}$
    \REPEAT
    \STATE currPin = $q$
    \STATE currSteps = SampleWalkLength($\alpha$)
    \FOR {i = [1 : currSteps]}
      \STATE currBoard = $E$(currPin)[rand()]
      \STATE currPin = $E$(currBoard)[randNeighbor()]
      \STATE $V$[currPin]++
    \ENDFOR
    \STATE totSteps += currSteps
    \UNTIL {totSteps  $\geq N$}
    \RETURN $V$
 \end{algorithmic}
\caption{Basic Random Walk; $q$ is the query pin; $E$ denotes the edges of graph $G$; $\alpha$ determines the length of walks; $N$ is the total number of steps of the walk; $V$ stores pin visit counts.}
\label{alg:rw}
\end{algorithm}

Having described the basic random walk procedure we now generalize and extend it to develop the Pixie Random Walk. Pixie Random Walk algorithm is comprised of Algorithms~\ref{alg:pixie_rw} and \ref{alg:pixie_rwn} and includes the following improvements over the basic random walk: (1) Biasing the random walk towards user-specific pins; (2) Multiple query pins each with a different weight; (3) Multi-hit booster that boosts pins that are related to multiple query pins; (4) Early stopping that minimizes the number of steps of the random walk while maintaining the quality of the results.

\hide{
\begin{algorithm}[t]
\vspace{-0pt}{\sc RandomWalkMultiple}($Q$: Set of query pins,
            $W$: Set of weights for query pins,
			      $E$: Set of edges,
			      $\alpha$: Real,
                              $N$: Int) \\
 \begin{algorithmic}[1]
    \FORALL {$q\in Q$}
      \STATE $N_q$ = $w_qN$ \label{alg:rwn:weight}
      \STATE $V_q$ = {\sc RandomWalk}($q$, $E$, $\alpha$, $N_q$) \label{alg:rwn:rw}
    \ENDFOR
    \FORALL {$\pin\in G$}
      \STATE $V[\pin]$ = $\sum_{q\in Q}V_q[p]$ \label{alg:rwn:sum}
    \ENDFOR
    \RETURN $V$
 \end{algorithmic}
\caption{Basic recommendations for multiple pins. $W$ is a set of weights of query pins $Q$}
\label{alg:rwn}
\end{algorithm}
} 

\xhdr{(1) Biasing the Pixie Random Walk}
\label{sec:local}
It is important to bias the random walk in a user-specific way. This way, even for the same query set $Q$, recommendations will be personalized and will differ from a user to user.
For example, Pinterest graph contains pins and boards with different languages and topics and from the user engagement point of view it is important that users receive recommendations in their language and on the topic of interest. 

We solve the problem of biasing the random walk by changing the random edge selection to be biased based on user features. The random walk then prefers to traverse edges that are more relevant to that user. One can think of these edges as having higher weight/importance than the rest of the edges in the graph. This way we bias the random walk in a user-specific way towards a particular part of the graph and let it focus on a particular subset of pins. In practice, this modification turns out to be very important as it improves personalization, quality, and topicality of recommendations, which then leads to higher user engagement.

Pixie algorithm takes as input a set of user features $U$ (Algorithm~\ref{alg:pixie_rw}). Notice that between different Pixie calls for different users and queries we can choose bias edge selection dynamically based on user and edge features which increases the flexibility of Pixie recommendations. 
In particular, {\sc PixieRandomWalk} selects edges with {\tt PersonalizedNeighbor(E,U)} to prefer edges important for user $U$. This allows us to prefer edges that match users features/preferences such as language or topic. Conceptually, this allows us to bias the walk in a user specific way but with minimal storage as well as computational overhead. Essentially, one could think of this method as using a different graph for each user where edge weights are tailored to that user (but without the need to store a different graph for each of the 200+ million users). In practice for performance reasons we currently limits the weights to only take values from a discrete set of possible values. We further avoid overhead by storing edges for similar languages and topics consecutively in memory and thus {\tt PersonalizedNeighbor(E,U)} is a subrange operator.

\xhdr{(2) Multiple Query Pins with Weights}
\label{sec:rec_multi_pins}
To holistically model the user it is important to make recommendations based on the entire historical context of a given user. We achieve this by performing queries based on multiple pins rather just on one pin. Each pin $q$ in the query set $Q$ is assigned a different weight $w_q$.
Weights are based on the time since the user interacted with a pin and the type of interaction.
To produce recommendations for a set $Q$ of query pins we proceed as follows. We run Pixie Random Walk (Algorithm~\ref{alg:pixie_rw}) from each query pin $q \in Q$ and maintain a separate counter $V_q[p]$ of pin $p$ for each query pin $q$. Last, we combine the visit counts by applying a novel formula which we describe later.

An important insight here is that the number of steps required to obtain meaningful visit counts depends on the query pin's degree. Recommending from a high-degree query pin that occurs in many boards requires many more steps than from a pin with a small degree. Hence, we scale the number of steps allocated to each query pin to be proportional to its degree. However, the challenge remains that if we assign the number of steps in linear proportion to the degree then we can end up allocating not even a single step to pins with low degrees.

We achieve our goal of step distribution by allocating the number of steps based on a function that increases sub-linearly with the query pin degree and scale the per pin weights $w_q$ by a scaling factor $s_q$. We construct the following scaling factor for each pin:
\begin{equation}
  s_q = |E(q)|\cdot\left(C-\log|E(q)|\right)
\end{equation}
where $s_q$ is the scaling factor for a query pin $q\in Q$, $|E(q)|$ is the degree of $q$,
and $C={\it max}_{p\in \Pins} |E(p)|$ is the maximum pin degree. This function, by design,  does not give disproportionately high weights to popular pins. We then allocate the number of steps as the following:
\begin{equation}
  N_q= w_q N \frac{s_q}{\sum_{r \in Q} {s_r}}
  \label{eq:pixieNq}
\end{equation}
where $N_q$ is the total number of steps assigned to the random walks that start from query pin $q$. The distribution gives us the desired property that more steps are allocated to starting pins with high degrees and pins with low degrees also receive sufficient number of steps. We implement this in line 2 of Algorithm~\ref{alg:pixie_rwn}.

\xhdr{(3) Multi-hit Booster}
\label{sec:multihit}
Another innovation of Pixie algorithm is that in general for queries with a set $Q$ of query pins, we prefer recommendations that are related to multiple query pins in $Q$. Intuitively, the more query pins a candidate pin is related to, the more relevant it is to the entire query. 
In other words, candidates with high visit counts from multiple query pins are more relevant to the query than for example candidates having equally high total visit count but all coming from a single query pin.

The insight here is that we let Pixie boost the scores of candidate pins that are visited from multiple query pins. We achieve this by aggregating visit counts $V_q[p]$ of a given pin $p$ in a novel way. Rather than simply summing visit counts $V_q[p]$ for a given pin $p$ over all the query pins $q\in Q$, we transform them and this way reward pins that get visited multiple times from multiple different query pins $q$: 
\begin{equation}
  V[\pin] = \left(\sum_{q\in Q} \sqrt{V_q[p]}\right)^2
\end{equation}
where $V[\pin]$ is the combined visit count for pin $\pin$. Note that when a candidate pin $\pin$ is visited by walks from only a single query pin $q$ then the count is unchanged. However, if the candidate pin is visited from multiple query pins then the count is boosted. Subsequently, when the top visited pins are selected from the counter $V$, the proportion of ``multi-hit'' pins is higher. We implement this in line 5 of Algorithm~\ref{alg:pixie_rwn}.

\xhdr{(4) Early Stopping}
\label{sec:earlystop}
The procedures described so far would run random walks for a given fixed number of steps $N_q$. However, since the Pixie runtime is critically dependent on the number of steps we want to run walks for smallest possible number of steps. Here we show that we can substantially reduce the runtime by adapting the number of steps $N_q$ depending on the query $q$ rather than having a fixed $N_q$ for all query pins.

Our solution to this problem is to terminate the walks once the set of top candidates becomes stable, \ie, does not change much with more steps. Since Pixie recommends thousands of pins, if implemented na\"{\i}vely, this monitoring can be more expensive than the random walk itself.  
However, our solution elegantly overcomes this by having two integers $n_p$ and $n_v$. We then terminate the walks when at least $n_p$ candidate pins have been visited at least $n_v$ times. This monitoring is easy and efficient to implement because we only need a counter to keep track of the number of candidate pins that have been visited at least $n_v$ times (lines 12-15 of Algorithm~\ref{alg:pixie_rw}).

We later show in Section~\ref{sec:eval_early_stop} that early stopping produces almost the same results as a long random walk but in about half the number of steps, which speeds-up the algorithm by a factor of two.

\begin{algorithm}[t]
\vspace{-0pt}{\sc PixieRandomWalk}($q$: Query pin,
            $E$: Set of edges,
            $U$: User personalization features,
            $\alpha$: Real,
            $N$: Int, $n_p$: Int, $n_v$: Int)\\
 \begin{algorithmic}[1]
    \STATE totSteps = 0, $V$ = $\vec{0}$
    \STATE nHighVisited = 0
    \REPEAT
    \STATE currPin = $q$
    \STATE currSteps = SampleWalkLength($\alpha$)
    \FOR {i = [1 : currSteps]}
      \STATE currBoard = $E$(currPin)[PersonalizedNeighbor(E,U)]
      \STATE currPin = $E$(currBoard)[PersonalizedNeighbor(E,U)]
      \STATE $V$[currPin]++
      \IF {$V$[currPin] == $n_v$}
             \STATE nHighVisited++
      \ENDIF
    \ENDFOR
    \STATE totSteps += currSteps
    \UNTIL {totSteps $\geq N$ or nHighVisited  $> n_p$}
    \RETURN $V$
 \end{algorithmic}
\caption{Pixie Random Walk algorithm with early stopping.}
\label{alg:pixie_rw}
\end{algorithm}

\begin{algorithm}[t]
\vspace{-0pt}{\sc PixieRandomWalkMultiple}($Q$: Query pins,
                              $W$: Set of weights for query pins,
            $E$: Set of edges,
            $U$: User personalization features,
            $\alpha$: Real, $N$: Int)\\
 \begin{algorithmic}[1]
    \FORALL {$q\in Q$}
      \STATE $N_q$ = Eq.~\ref{eq:pixieNq} \label{alg:rwn:weight}
      \STATE $V_q$ = {\sc PixieRandomWalk}($q$, $E$, $U$, $\alpha$, $N_q$) \label{alg:rwn:rw}
    \ENDFOR
    \FORALL {$\pin\in G$}
      \STATE 
      $V[\pin]$ = $\left(\sum_{q\in Q} \sqrt{V_q[p]}\right)^2$ \label{alg:rwn:sum}
    \ENDFOR
    \RETURN $V$
 \end{algorithmic}
\caption{Pixie recommendations for multiple pins.}
\label{alg:pixie_rwn}
\end{algorithm}

\subsection{Graph Pruning}
\label{sec:graph_curation}

Another important innovation of our method is graph cleaning and pruning. 
Graph pruning improves recommendation quality and also decreases the size of the graph so that it can fit on a smaller, cheaper machine with better cache performance for serving.

The original Pinterest graph has 7 billion nodes and over 100 billion edges. However, not all boards on Pinterest are topically focused. Large diverse boards diffuse the walk in too many directions, which then leads to low recommendation performance. Similarly, many pins are mis-categorized into wrong boards. The graph prunning procedure cleans the graph and makes it more topically focused.
As a side benefit graph pruning also leads to a much smaller graph that fits into the main memory of a single machine.  Not having to distribute the graph across multiple machines, leads to huge performance benefits because the random walk does not have to ``jump'' across the machines. 

We approach the problem of graph prunning as follows. First, we quantify the content diversity of each board by computing the entropy of its topic distribution. 
We run LDA topic models on each pin description to obtain probabilistic topic vectors. We then use topic vectors of the latest pins added to a board as input signals to compute board entropy. Boards with large entropy are removed from the graph along with all their edges. 

Another challenge is that real-world graphs have skewed heavy-tailed degree distributions. In the case of Pinterest this means that some pins are extremely popular and have been saved to millions of boards. For such nodes, the random walk needs to be run for many steps because it gets diffused among a large number of network neighbors. We address this problem by systematically discarding edges of high degree pins. However, rather than discarding the edges randomly we discard edges where a pin is miscategorized and does not belong topically into a board.
We use the same topic vectors as above and calculate the similarity of a pin to a board using the cosine similarity of the topic vectors and then only keep the edges with the highest cosine similarity. In particular, the extent of pruning is determined by a {\it pruning factor} $\delta$. 
We update the degree of each pin $\pin$ to $|E(\pin)|^\delta$ and discard the edges that connect $\pin$ to boards whose topic vectors have low cosine similarity with the topic vector of $\pin$. 

After pruning the graph contains 1 billion boards, 2 billion pins and 17 billion edges. Surprisingly, we find that pruning is beneficial in two aspects: (1) decreases the size of the graph (and the memory footprint) by a factor of six; (2) and also leads to 58\% more relevant recommendations (further details are described in Section~\ref{sec:eval_graph}).

\subsection{Implementation}
\label{sec:impl}

Here we shall discuss the implementation details of Pixie. To meet the real-time requirements Pixie relies on efficient data structures which we discuss first. Then we discuss how the graph of pins and boards is generated. Finally, we briefly discuss the servers that respond to Pixie queries constructed by various applications at Pinterest. Pixie is written in C++ and is built on top of the Stanford Network Analysis Platform (SNAP)~\cite{snap}.

\xhdr{Graph Data Structure}
The random walk procedure described in Algorithm~\ref{alg:pixie_rw} spends most of its time in the inner loop (lines 6 to 13). Therefore, for this procedure to be effective, we need efficient implementations of graphs and counters. We describe these next.

Consider lines 7-10 of Algorithm~\ref{alg:pixie_rw}. For these operations to be efficient we need to quickly sample a board connected to a pin and a pin connected to a board. We next design a data structure that performs these operations in constant time.

We develop a custom highly optimized data structure where we assign each node of $G$, \ie, every pin $\pin\in\Pins$ and board $b\in B$ a unique ID between $1$ and $|\Pins\cup\Boards|$.
The graph is implemented as a variant of the standard adjacency list implementation. Each node is associated with a list of its neighbors. Allocating each such list dynamically is slow and causes memory fragmentation, therefore we use the object pool pattern to concatenate all of the adjacency lists together in one contiguous array {\it edgeVec}.

In {\it edgeVec} the $i^{th}$
entry ${\it offset}_i$ is the offset where node $i$'s neighbors are stored in the associated {\it edgeVec}. Note that the number of node $i$'s neighbors is given by $\it{offset}_{i+1} - \it{offset}_{i}$.
To sample a neighbor of a node with ID $i$ whose neighbors are stored in an {\it edgeVec} $F$, we  read the ID stored at
  \begin{equation}
  F\left[{\it offset}_i+\left(rand()\%({\it offset}_{i+1}-{\it offset}_i)\right)\right]
  \end{equation}
Thus, the accesses on lines 5, 8, and 10 of Algorithm~\ref{alg:pixie_rw} can be performed efficiently in constant time.

\xhdr{Visit Counter} 
After sampling a pin $\pin$, {\sc PixieRandomWalk} increments the visit count---the
number of times $\pin$ has been visited (line 11, Algorithm~\ref{alg:pixie_rw}).
We develop an open addressing hash table $V$ that implements this operation efficiently with linear probing. First, we allocate a fixed size array where each element is a key-value pair. For Pixie, the keys are pin IDs in $G$ and the values are visit counts.
When incrementing the visit count of a pin ID $k$, we first use a hash function (described below) to index into this array.  If $k$ matches the key stored at the index then we update the value. Otherwise, we continue probing the following indices until we either find a free element or $k$ (\ie, linear probing).
In the former case, we assign the key $k$ and value 1 to the free element.
Our primary motivation to use linear probing is to maintain good cache locality while resolving collisions.
For this procedure to be efficient, the hash function needs to be fast. We use the very light-weight multiplicative hash function (\ie, multiply the key with a fixed prime number modulo the array size).
Empirically, we have observed that the key insertions in this counter have performance comparable to random array accesses.  After {\sc PixieRandomWalkMultiple} terminates, the array is sorted in descending order of values and the pin IDs with top visit counts are returned as recommendations.

Using arrays to implement hash tables can have problems; for example, if the array is full then it would need to be resized. In the context of Pixie, the number of steps $N$ provides an upper bound on the number of keys the hash table needs to support as the number of pins with non-zero visit counts can never exceed the number of steps. We conservatively allocate an array of size $N$ when initializing the hash table to avoid resizing.

\xhdr{Graph Generation and Pruning}
\label{sec:graphupdate}
The graph generation first runs a Hadoop pipeline, followed by a graph compiler.
The Hadoop pipeline contains a series of MapReduce jobs that go through the data at Pinterest
and retrieve all the boards and the pins belonging to them. The pipeline outputs a raw text file that contains the edges between boards and pins, and uploads it to a global storage. The graph compiler runs on a single terabyte-scale RAM machine that polls for new raw graphs. Once a new raw graph file is available, the graph compiler downloads and parses the raw data into memory, prunes the graph, then persists it to disk in a binary format. These binaries can be shared easily between machines. This graph generation process runs once per day.

Loading the graph binaries from the disk to shared memory takes $\approx$10 minutes. The process is efficient as the load is a sequential read from the disk. 
We use Linux HugePages for this shared memory to increase the size of each virtual memory page from 4 KB to 2 MB thus decreasing the number of page table entries needed by a factor 512. Too many page table entries is especially problematic on virtual machines; the HugePages option enabled Pixie on virtual machines to serve twice as many requests at half the runtime.

\xhdr{Pixie Server}
\label{sec:server}
On start up, each Pixie server loads the graph from the disk into memory.
Each server has an IO thread pool and a worker thread pool. The IO threads serialize and deserialize queries and responses. They hand off sets of pins to worker threads. Each worker thread has its own counter that collects visit counts.
To avoid synchronization costs among the workers,
each query is served by a single worker. The server also has a background thread that periodically checks for
the availability of new graphs. When available, the latest graph is downloaded to the disk. The server restarts once a day and loads the latest available graph in memory.

\section{Experiments}
\label{sec:experiments}

In this section we evaluate Pixie and empirically validate its performance. We quantify the quality of Pixie recommendations, the runtime performance of the Pixie algorithm, and the effect of the graph pruning on recommendations.

\subsection{Pixie Recommendation Quality}

\begin{table}
\begin{tabular}{c | c | c | c}
\hline
Method & Top 10 & Top 100 & Top 1000  \\
\hline
Content-based (textual)  & 1.0\% & 2.2\% & 4.8\%  \\
Content-based (visual)  & 1.1\% & 2.4\%  & 4.5\% \\
Content-based (combined)  & 2.1\% & 4.6\%  & 10.5\% \\
Pixie (graph-based) & 6.3\% & 23.1\% & 52.2\% \\
\hline
\end{tabular}
\caption{Given a query pin, predict which pin will be repinned.
Performance is quantified by fraction of times the correct pin was ranked among top $K$.}
\vspace{-8mm}
\label{tab:pixie_topk}
\end{table}

The goal of Pixe is to produce highly-engaging recommendations. We quantify the quality of recommendations in two ways: (1) given a user we aim to predict which pin they will engage with next; (2) we perform A/B experiments where we can directly measure the lift in user engagement due to recommendations made by Pixie.

For comparison we use two state-of-the-art deep learning content-based recommender methods
that use visual and textual features of a given pin to produce recommendations. Each pin is associated with both an image and a set of textual annotations defined by the users. We use these content features to create embeddings of pins. To generate recommendations for a given query pin $q$ we then apply nearest neighbor methods to find the most similar pins.

The visual embeddings are the 6-th fully connected layer of a classification network using the VGG-16 architecture \cite{simonyan2014very,donahue2014decaf}.
The textual annotation embeddings are trained using the Word2Vec model \cite{mikolov2013distributed}. The context of an annotation consists of annotations that associate with each pin.
For generating recommendations based on visual embeddings we use the hamming distance, while for generating recommendations based on textual embeddings we use the cosine distance.

\xhdr{Ranking the most related pin}
We quantify the success of recommendations by performing the following prediction task: Given a user that is examining a query pin $q$, we aim to predict which of all other pins is most related to the query pin. Here we rely on user activity and say that pin $x$ is most related to $q$ if the user while looking at $q$ has saved its related pin $x$. 
We formulate this as a ranking task where given a query pin $q$ we aim to rank all other 2B pins with the goal to rank the saved pin $x$ as high as possible. We measure the performance by {\em hit rate}, which we define as the fraction of times the saved pin was ranked among the top-$K$ results.

Table~\ref{tab:pixie_topk} gives the hit rate at $K=$ 10, 100, and 1000. Pixie gives much better performance of recommendations and is able to predict which of the pins is most related to the query pin (and thus saved by the user) most accurately for all values of $K$. Note that due to large-scale nature of our recommendation problem we do not compare to other baselines (such as collaborative filtering or matrix factorization methods) since no ready-to-use methods exist.

\xhdr{Results of A/B experiments}
The ultimate test of the quality of Pixie recommendations is the lift in user engagement in a controlled A/B experiment where a random set of users experiences Pixie recommendations, while other users experience recommendations given by the old Hadoo-based production system that precomputes recommendations and is not real-time. Any difference in engagement between these two groups can be attributed the increased quality of Pixie recommendations. We measure engagement by quantifying the fraction of pins that a user engages in by clicking, liking, or saving them. Table~\ref{tab:pixie_ab} summarizes the lifts in engagement of pins recommended by Pixie in controlled A/B experiments with observed increases between 13\% to 48\%.

\begin{table}
\begin{tabular}{c | c }
\hline
Experiment & Lift \\
\hline
Homefeed, per pin engagement  
& +48\% \\ 
Related pins, per pin engagement & +13\% \\ 
Board recommendations, per pin engagement  & +26\% \\ 
Localization, pins in user local language & +48-75\% \\ 
Explore tab, per pin engagement & +20\% \\ 
\hline
\end{tabular}
\caption{Summary of A/B experiments across different Pinterest
user surfaces. Lift in engagement of Pixie vs. current production systems.}
\vspace{-8mm}
\label{tab:pixie_ab}
\end{table}

\subsection{Pixie Performance and Stability}

Next we shall evaluate the runtime performance of Pixie algorithm.

\xhdr{Pixie Runtime}
\label{sec:rw_speed}
We evaluate how the number of steps and the size of the query set $Q$ affect the runtime of Pixie. For the experiment we sample 20,000 queries of size 1 and compute the average runtime for each $N$. Figure~\ref{fig:latency_num_steps}(a) shows that the runtime increases linearly with the number of steps. Moreover, the runtime is below 50 milliseconds for random walks with less than 200,000 steps.

\hide{
\begin{figure}[t]
     \centering
     \includegraphics[width=0.5\textwidth]{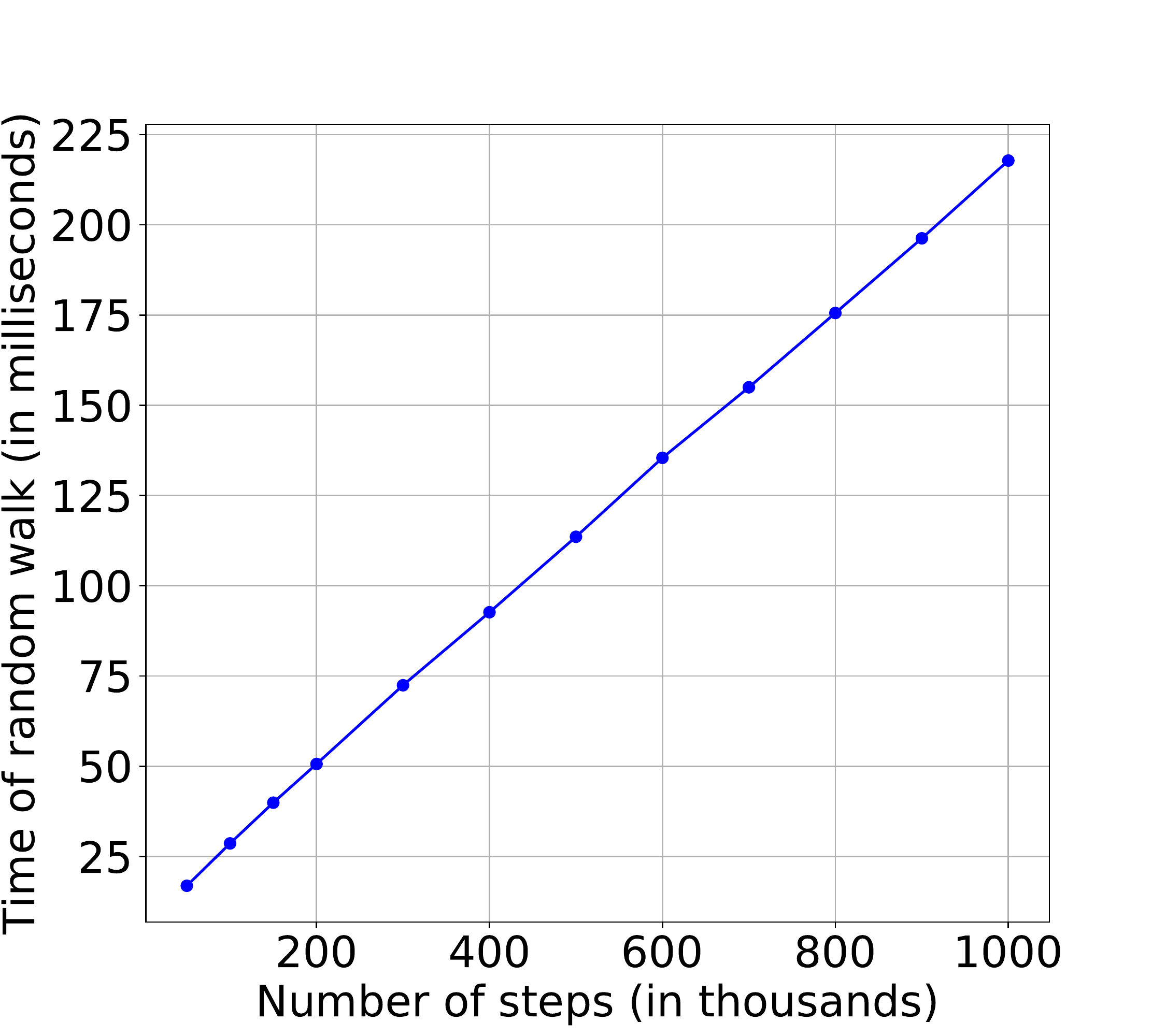}
     \caption{The runtime of {\sc RandomWalk} against the number of steps.}
     \label{fig:latency_num_steps}
\end{figure}
\begin{figure}[t]
     \centering
     \includegraphics[width=0.5\textwidth]{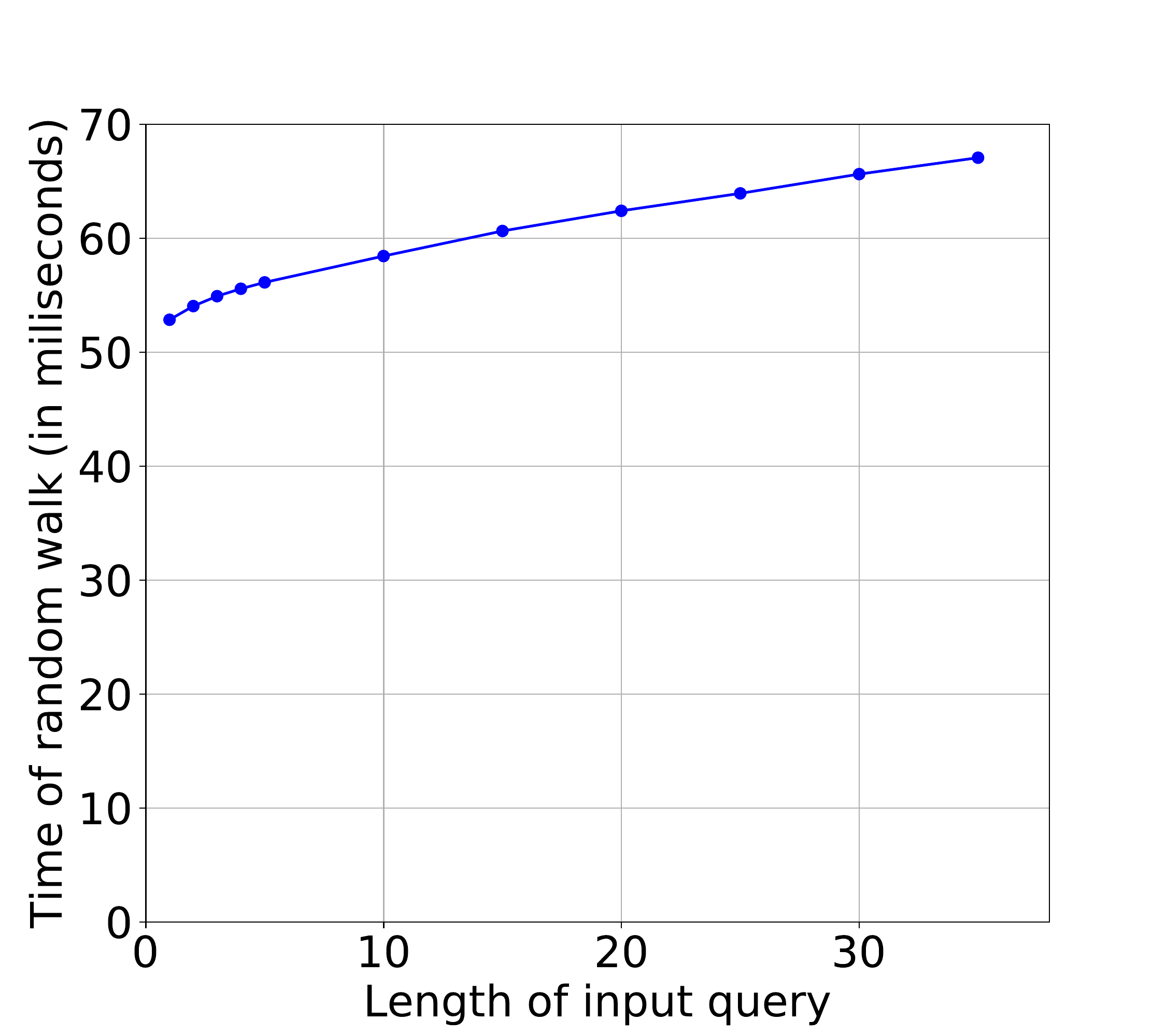}
     \caption{The runtime of {\sc RandomWalkMultiple} against the query size.}
     \label{fig:latency_query_lens}
\end{figure}
}

\begin{figure}[t]
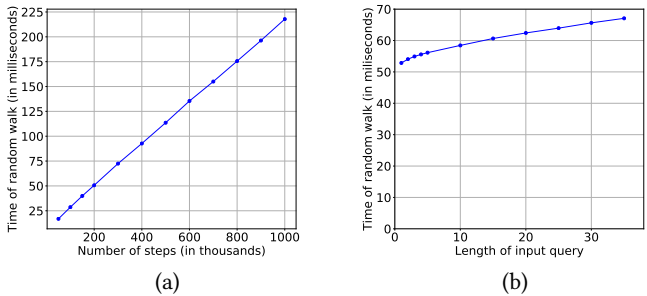

     \centering
     \begin{tabular}{cc}
     \includegraphics[width=0.24\textwidth]{FIG/latency_num_steps} &
     \includegraphics[width=0.24\textwidth]{FIG/latency_query_lens} \\
     (a) & (b)
     \end{tabular}
     \vspace{-3mm}
     \caption{(a) Runtime of {\sc PixieRandomWalk} against the number of steps and (b) against the size of the query set.}
     \label{fig:latency_num_steps}
     \label{fig:latency_query_lens}
     \vspace{-3mm}
\end{figure}

To evaluate runtime as a function of the query size, we randomly sample 20,000 queries of each query size. We keep the number of steps constant and compute the average runtime for queries with identical size. Figure~\ref{fig:latency_query_lens}(b) shows that the runtime increases slowly with the query size. This increase is primarily due to cache misses. During random walks from a query pin, the cache gets warmed with the neighborhood around the pin. When the walks are started from a different query pin, the cache becomes cold and needs to be warmed again. With longer queries, the cache becomes cold more often, and the runtime slightly increases.

\xhdr{Variance of Top Results}
\label{sec:rw_convergence}
The recommendations produced by a randomized procedure such as Pixie are not deterministic.
Each time Pixie is run, the visit counts depend on the sampled random numbers.
However, we desire {\it stability} of recommendations, \ie, the set of recommended pins
should not change much across multiple runs. If we could run random walks for a enough steps for the walk to converge then  the recommendations would become stable. However, Pixie has a tight runtime requirement and we cannot run random walks for billions of steps.
We study how the stability of the set of top visited pins
 varies with the number of steps with the aim to balance the conflicting requirements of low runtime and high stability.

\begin{figure}[t]
     \centering
     \vspace{-4.5mm}
     \includegraphics[width=0.45\textwidth]{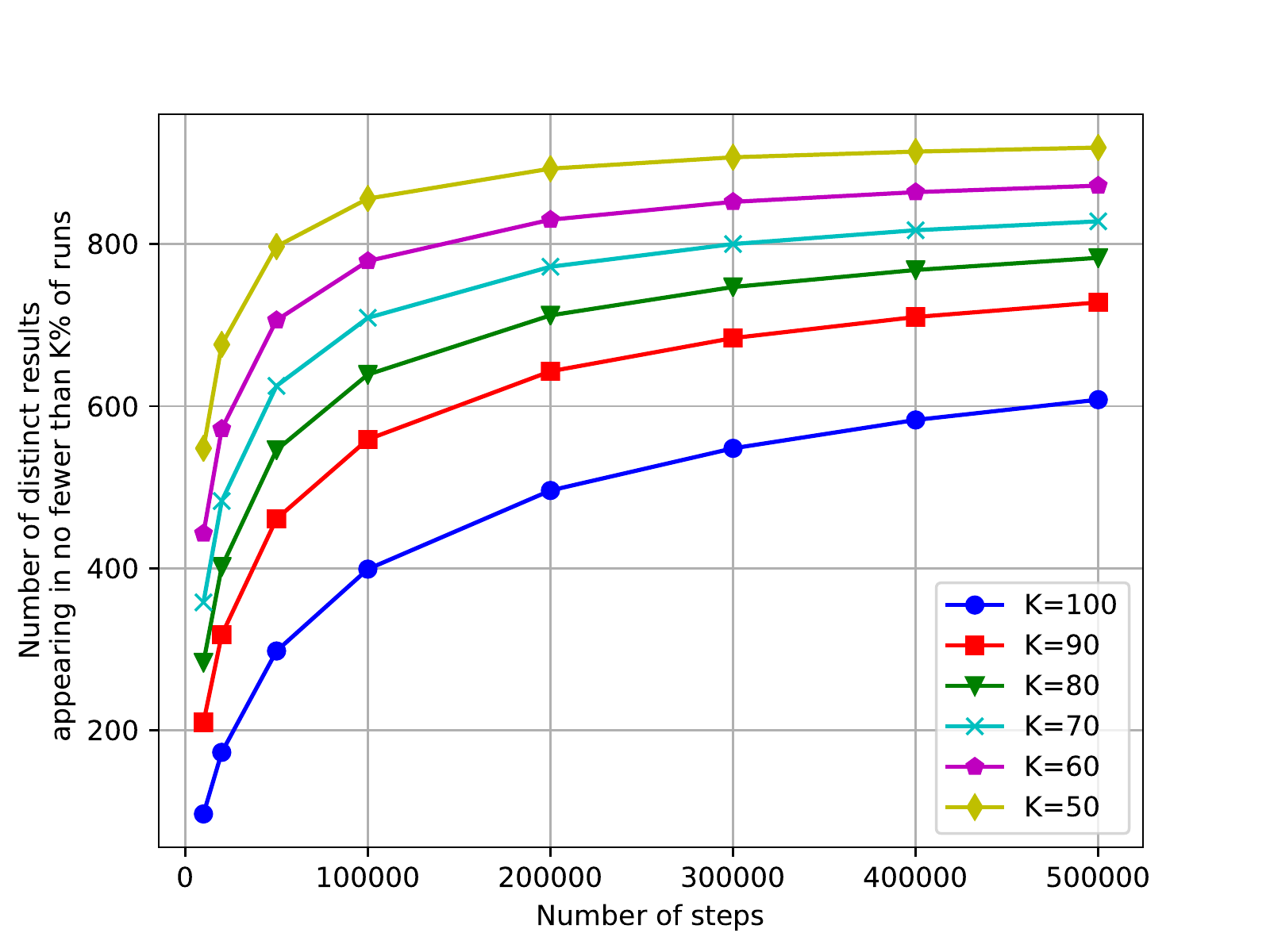}
     \vspace{-3mm}
     \caption{The variance of results against number of steps.}
     \label{fig:variance_num_steps}
     \vspace{-3mm}
\end{figure}

We randomly sample 20,000 queries of size 1 and then run each query 100 times. We then examine the top 1,000 results of each of 100 responses and count the number of pins which appear in at least $K$ responses ($K=50, 60, \ldots, 100$). Figure~\ref{fig:variance_num_steps} shows that the stability improves with the number of steps. With 100,000 steps, 400 results appear in all the responses and with 500,000 steps
600 results appear in all the responses ($K=100$). Moreover, the results are quite stable. For instance, with 100,000 steps,
more than 800 results appear in at least 50\% of the responses ($K=50$). Finally, the incremental gains in stability
obtained by increasing the number of steps declines after around 200,000 steps. Therefore, we conclude that several hundred thousand steps are
enough to obtain a stable recommendation set and increasing the number of steps beyond it does not help much.

\xhdr{Evaluation of the Biased Walk}
\label{sec:local_eval}
Here we evaluate the efficacy of biasing the Pixie Random Walk in a user-specific way. 
This way, even for the same query set, recommendations will be more personalized and will differ from a user to user.

To illustrate the effectiveness of the biasing procedure we consider the following experiment where the goal is to provide recommendations in a given target language. We start the random walk at a pin in some language 
and then aim to bias the Pixie Random Walk to visit pins from the target language.

We report the results in Table~\ref{tab:local_walk}. We consider three target languages: Japanese, Spanish, and Slovak. For each language, we show the percentage of target-languege pins in the response produced by simple random walks (Algorithm~\ref{alg:rw}) and Pixie Random Walk (Algorithm~\ref{alg:pixie_rw}). We consider two scenarios, when the query pins are in English (column 2) and when they are in target language (column 3). For queries originating from different languages, target language recommendations provide much better and more engaging user experience. We observe that Pixie Random Walk significantly boosts the target-language content in the query responses. 

\begin{table}[t]
\centering
\begin{tabular}{c|c|c} \hline
& En$\rightarrow$Japanese & Japanese$\rightarrow$Japanese  \\ \hline
{\sc BasicRandomWalk} & $16.35\%$ & $52.95\%$  \\ 
{\sc PixieRandomWalk} & $80.33\%$ & $100.00\%$ \\ \hline 
& En$\rightarrow$Spanish & Spanish$\rightarrow$Spanish  \\ \hline
{\sc BasicRandomWalk} & $41.94\%$ & $74.02\%$ \\ 
{\sc PixieRandomWalk} & $94.51\%$ & $100.00\%$ \\ \hline
& En$\rightarrow$Slovak & Slovak$\rightarrow$Slovak  \\ \hline
{\sc BasicRandomWalk} & $2.13\%$ & $16.06\%$ \\ 
{\sc PixieRandomWalk} & $42.55\%$ & $100.00\%$\\ \hline
\end{tabular}
\caption{Comparison of the proportion of target-language content produced by {\sc BasicRandomWalk} and {\sc PixieRandomWalk}. The second column shows the percentage of candidates in the target language when the query pin is in the English language and the third column shows the percentage when the query pin itself is in the target language.}
 \label{tab:local_walk}
 \vspace{-8mm}
\end{table}

\begin{figure*}[t]
     \vspace{-2mm}
     \centering
     \begin{tabular}{cc}
     \includegraphics[width=0.45\textwidth]{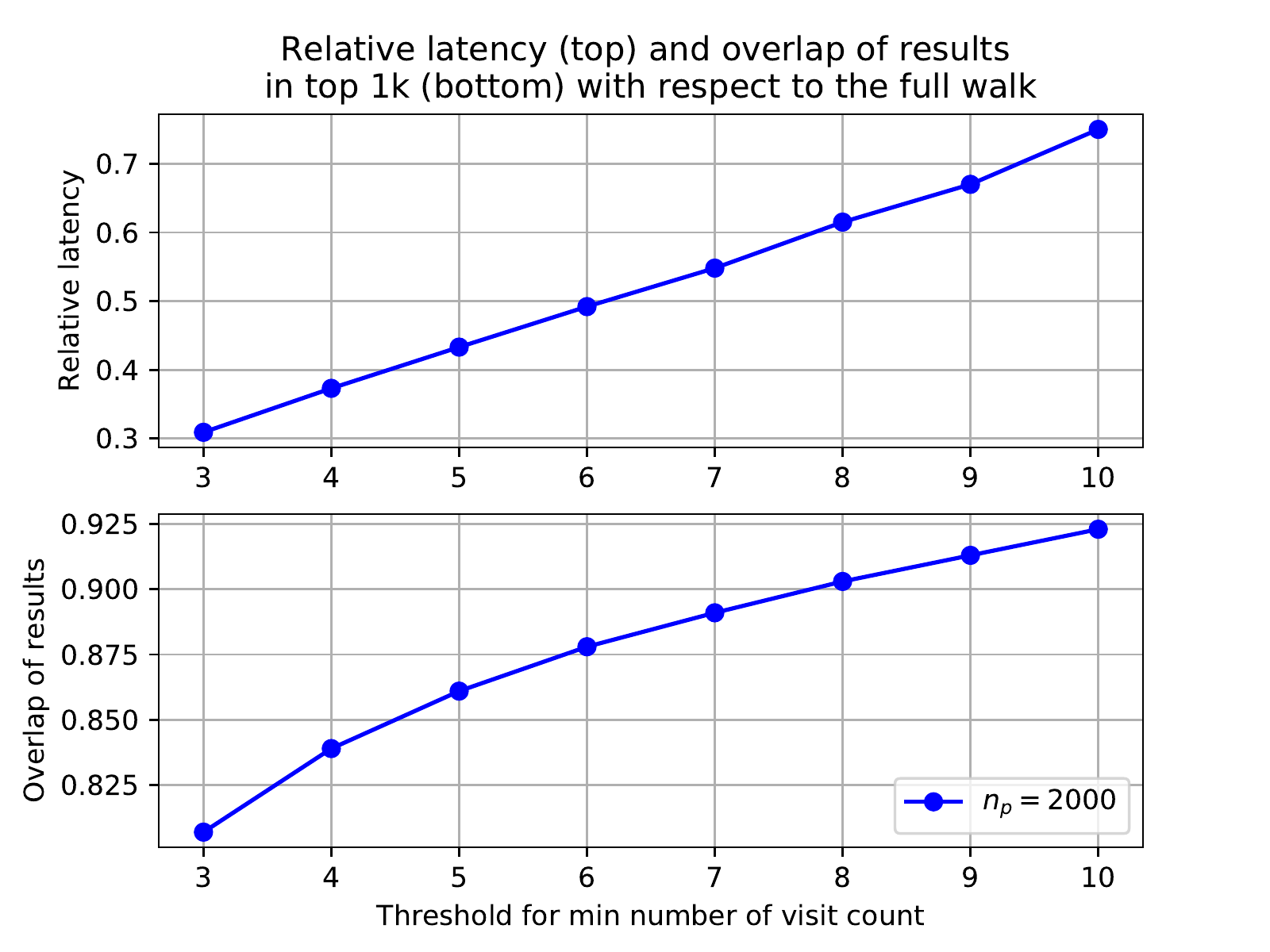} &
     \includegraphics[width=0.45\textwidth]{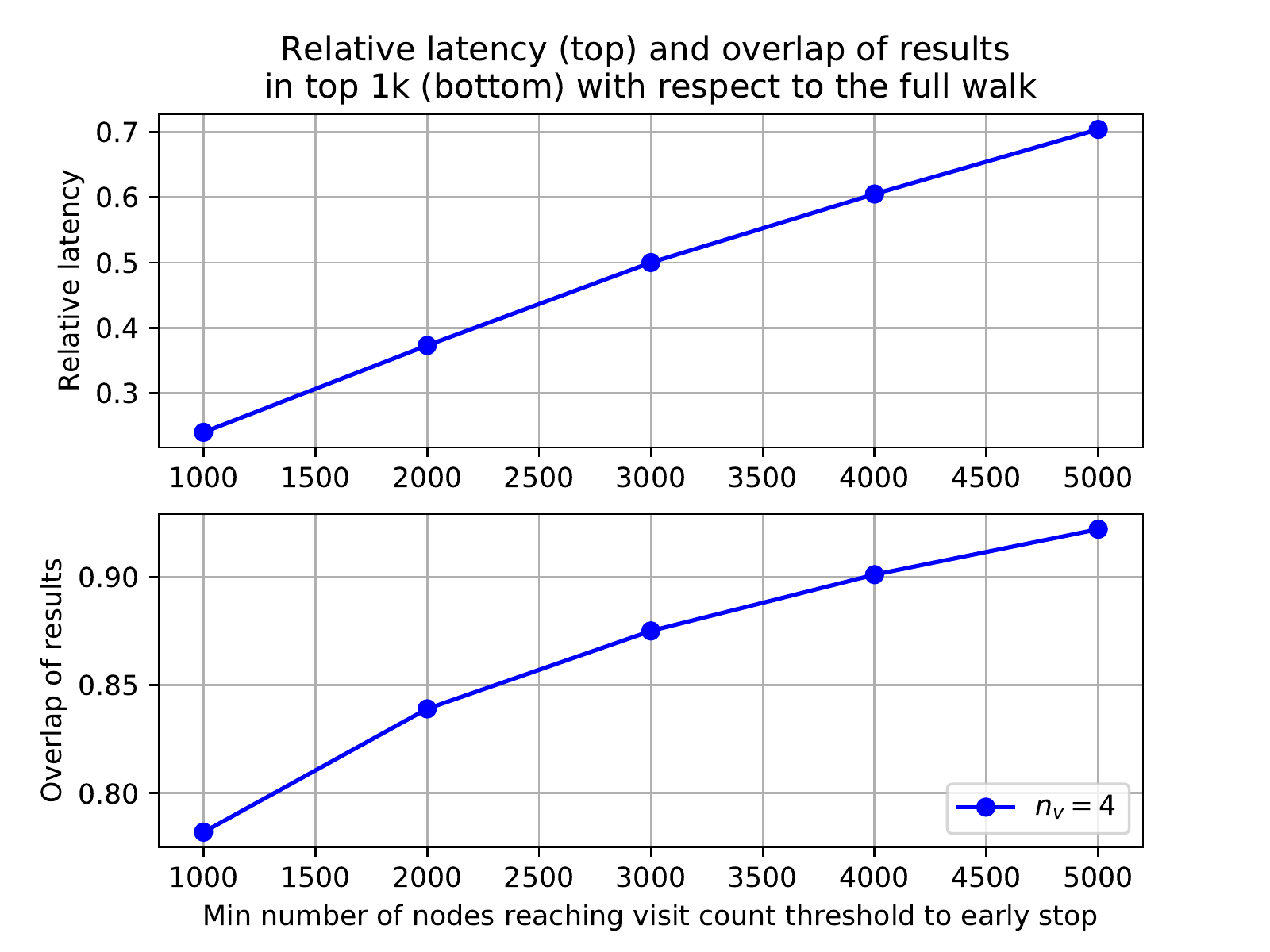}\\
     (a) & (b)
     \end{tabular}
     \vspace{-3mm}
     \caption{(a) Early stopping performance against $n_v$ with $n_p=2,000$. 
     (b) Early stopping performance against $n_p$  with $n_v=4$.
     }
     \label{fig:early_stop_vc}
     \label{fig:early_stop_num_nodes}
\end{figure*}

\xhdr{Early Stopping}
\label{sec:eval_early_stop}
Pixie algorithm terminates random walks after $n_p$ pins reach a visit count of $n_v$.
Thus $n_v$ acts as a minimum threshold of visit counts that the pins in the recommendation set have to meet
and $n_p$ denotes the minimum number of pins that reach this threshold before we stop the walks.
Lower values of $n_p$ and $n_v$ lead to lower running time but potentially also unstable recommendation results. 

Here we study how the runtime and stability is affected by $n_p$ and $n_v$ in order to set these parameters appropriately. We randomly sample 20,000 queries with size 1 and consider the top 1,000 recommendations. As a gold-standard set of recommendations we also run Pixie but with a fixed very large number of steps.

First, we set $n_p=2,000$ and vary $n_v$ with results shown in Figure~\ref{fig:early_stop_vc}(a). We observe that lower values of $n_v$ lead to much faster run times (Figure~\ref{fig:early_stop_vc}(a), top). For $n_v=6$, the run time reduces to half. With high values of $n_v$, we observe that the recommendations produced by Pixie have a high overlap with the gold-standard set. For example, for $n_v=8$, 900 (out of 1,000) results are common to both the recommendation sets, while the running time is improved by a factor of two.
Second, we fix $n_v=4$ and vary $n_p$ as shown in Figure~\ref{fig:early_stop_num_nodes}(b). The trends are similar. Overall, we observe that by choosing parameters appropriately, \eg, $n_p=2,000$ and $n_v=4$, we  achieve a good similarity with the gold-standard set of results (84\%), while improving the runtime for a factor of three.

\hide{
\begin{figure}[t]
     \vspace{-2mm}
     \centering
     \includegraphics[width=0.5\textwidth]{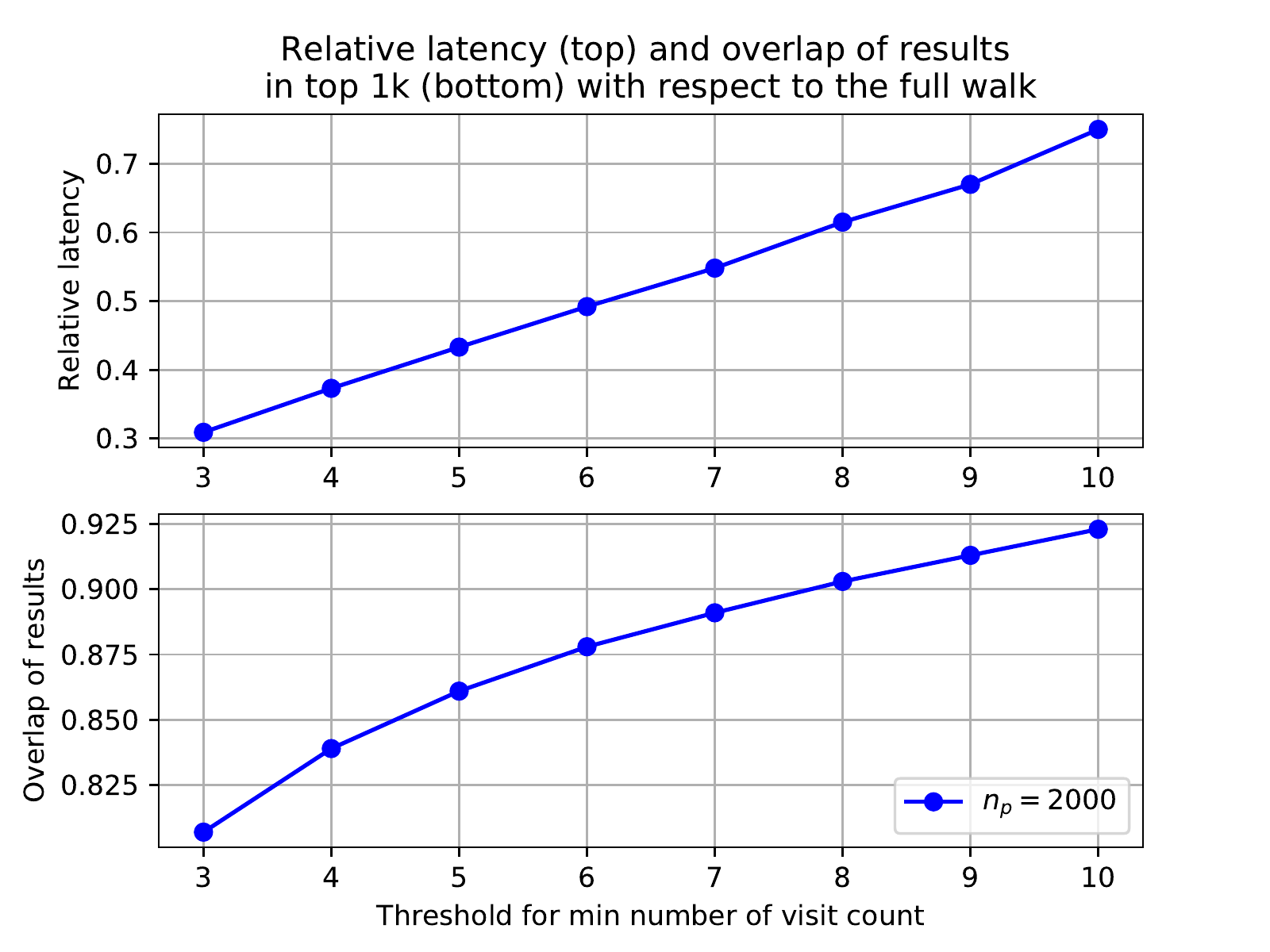}
     \caption{Early stopping performance with $n_p=2,000$.}
     \label{fig:early_stop_vc}
\end{figure}

\begin{figure}[t]
     \vspace{}
     \centering
     \includegraphics[width=0.5\textwidth]{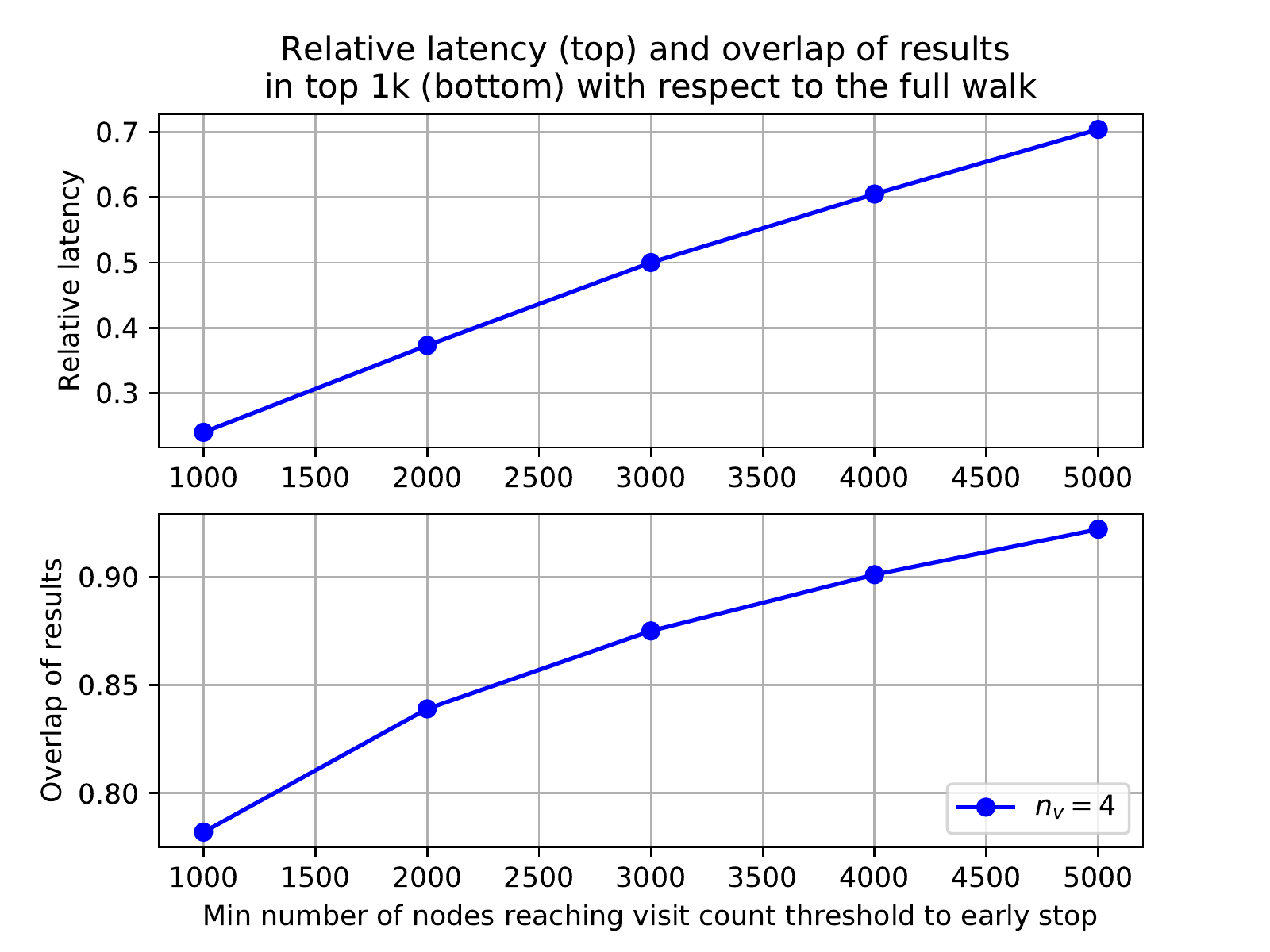}
     \caption{Early stopping performance against $n_p$  with $n_v=4$.}
     \label{fig:early_stop_num_nodes}
\end{figure}
} 

\subsection{Evaluating Graph Pruning}
\label{sec:eval_graph}

We evaluate the effect of graph pruning on the quality of Pixie recommendations, the memory usage, as well as running time.
We evaluate the quality on the {\it link prediction} task that we describe next. When a user adds a pin to a board they create a new edge in the graph. We use Pixie to predict the pins that would be saved to
a board after a timestamp $t$, denoted by $X$, by querying the pins that already exist on the board before $t$, denoted by $Q$.
Pixie succeeds if the response $R$ is identical to $X$. As is standard, we are interested in two  quantities: {\it recall} and {\it precision}. The recall measures the percentage of pins in $X$ that $R$ contains as well. And precision measures the percentage of pins in $R$ that are included in $X$.  We then use the F1 score of precision and recall as a measure of the quality of the results.

In this evaluation, we first select 100,000 boards at random. Then we take the latest 20 pins in each board before the time $t$. Each such sample constitutes a Pixie query $Q$. We select the top hundred visited pins as the recommendation set $R$. Finally, we compute the $F_1$ score using the pins added to the boards after time $t$ as $X$.

\begin{figure}[t]
    \vspace{-4mm}
    \centering
    \includegraphics[width=0.49\textwidth]{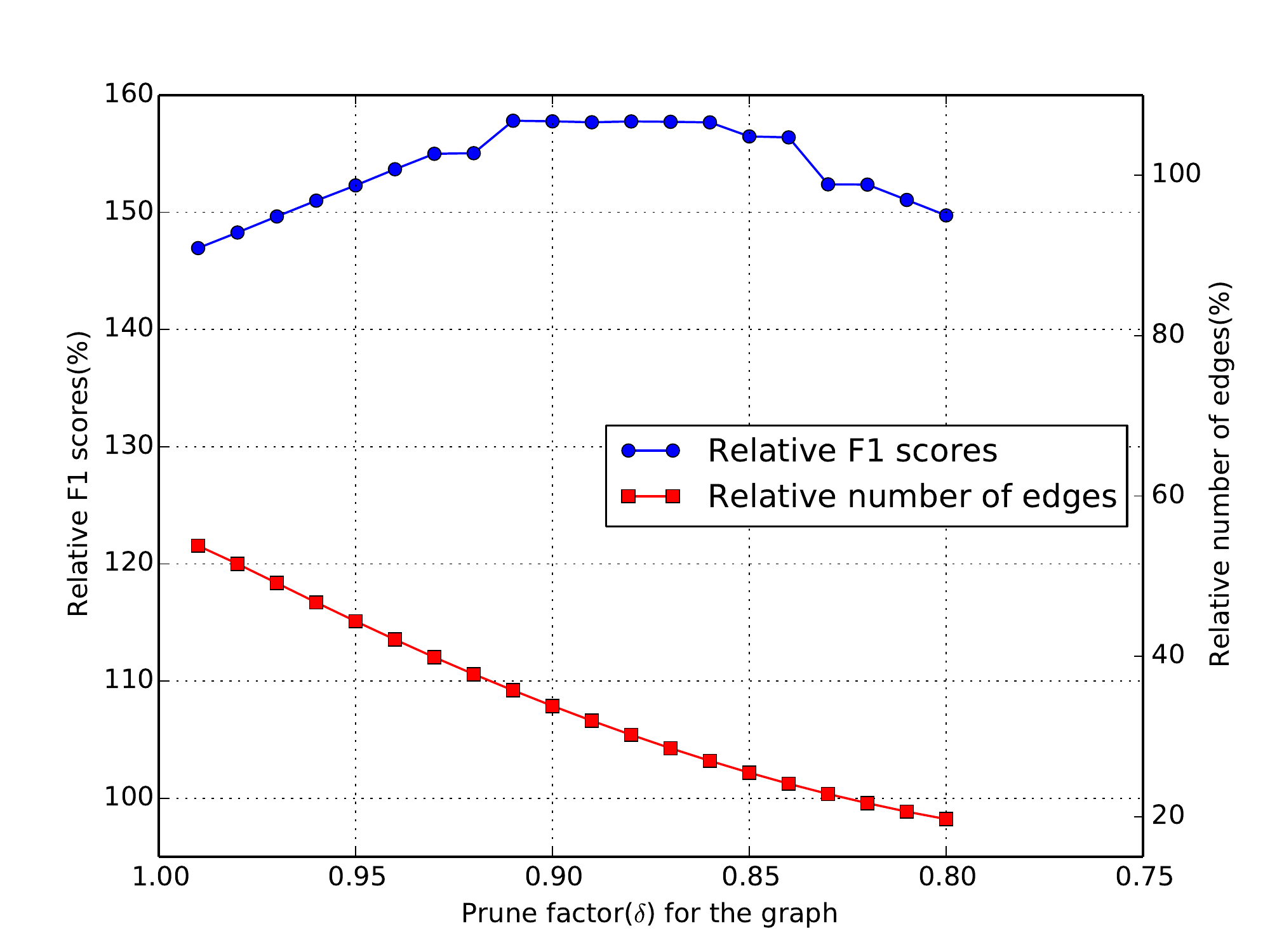}
    \caption{F1 scores for link prediction and number of edges for different graph pruning factors.}
    \label{fig:pin_to_pin_f1}
\end{figure}

We examine the effects of pruning by first removing the most diverse 10\% of boards as described in Section~\ref{sec:graph_curation}. We then examine the effect of pruning pin degree by varying the pruning factor $\delta$. Recall that $\delta = 1$ represents the full graph and decreasing $\delta$ prunes the graph more. Figure~\ref{fig:pin_to_pin_f1} shows that the number of edges in the pruned graph decreases with $\delta$ monotonically. 

The $F_1$ score changes as we keep pruning the graph. When $\delta$ becomes too low, even relevant edges are pruned and the quality of the recommendations deteriorates.
However, we also observe that graph pruning significantly improves the quality of recommendations. Figure~\ref{fig:pin_to_pin_f1} shows that when $\delta=0.91$, the $F_1$ score peaks at $58\%$ above the unpruned graph $F_1$ and the graph contains only $20\%$ the original number of edges. This means that the graph pruning actually improves the quality of recommendations by 58\%, while also pruning the graph to only a sixth of its size.

\begin{figure}[t]
     \centering
     \includegraphics[width=0.47\textwidth]{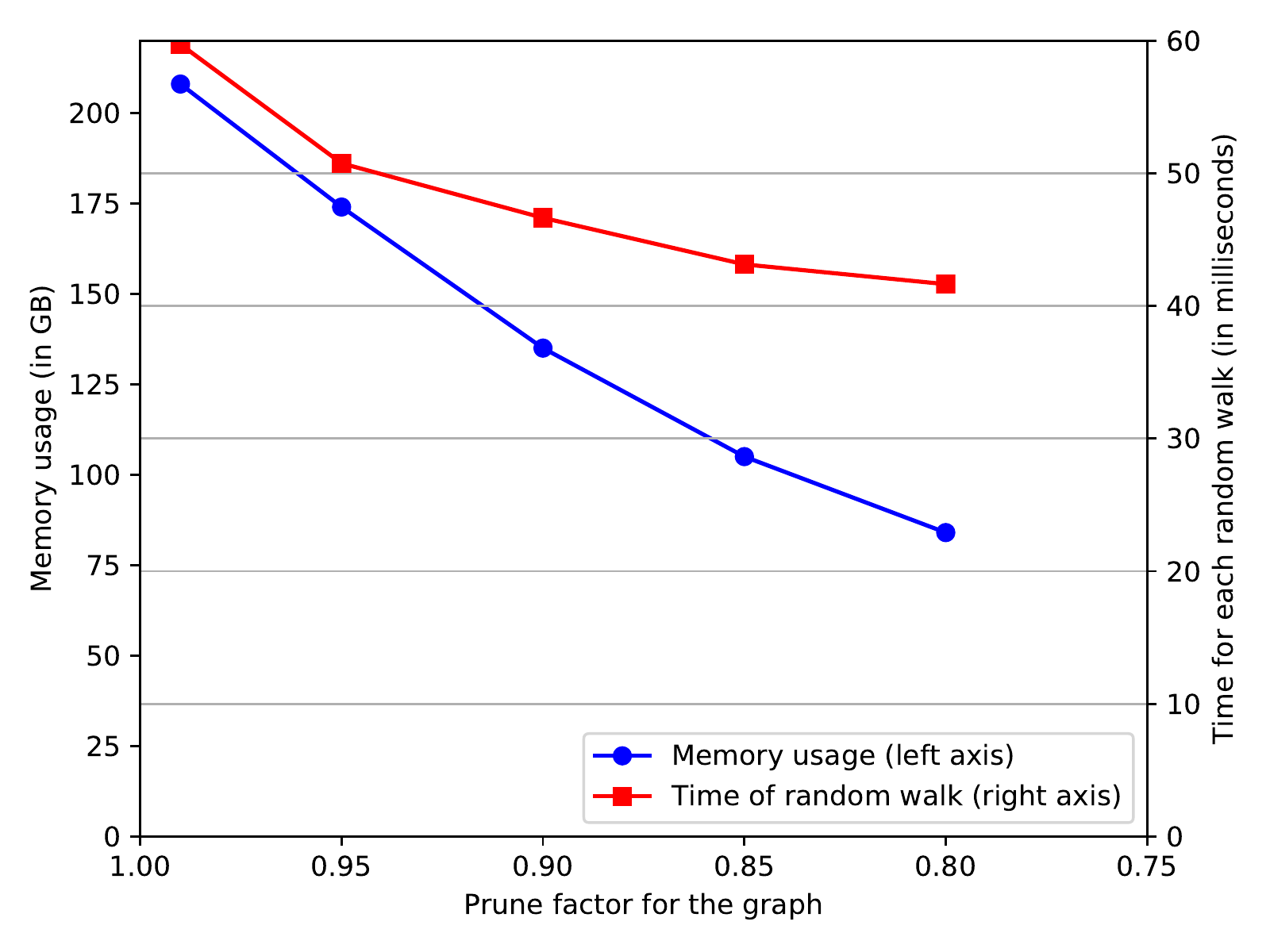}
     \caption{The memory usage and Pixie runtime against different pruned graphs.}
     \label{fig:memory_latency}
\end{figure}

Finally, we show how graph pruning affects the memory usage and the runtime of Pixie in Figure~\ref{fig:memory_latency}. As the size of the graph decreases, both the memory as well as the Pixie Random Walk runtime decrease significantly.

\section{Use cases at Pinterest}
\label{sec:discussion}

There are many Pinterest applications that use Pixie to generate
relevant and timely recommendations. We shall discuss some of them below.

\subsection{Homefeed}
When a user loads Pinterest, they view a grid of pins that they might find relevant to save in their boards. This grid is called the user Homefeed.
Using the real-time recommendations of Pixie, we are able to create an engaging and responsive Homefeed.
Every time the user takes an {\em action} on a pin (such as clicking, liking, or saving a pin),
we create a query, send it to a Pixie server, and refresh the recommendations.
More precisely, to obtain a Pixie query, we collect all pins where user performed an action and assign a weight to each pin. A single pin's initial weight depends on the action type, and decays with a half life of $\lambda$.
User's pins are then collected in to a single Pixie query.
The pin recommendations in the Pixie response are ranked and added to the user's Homefeed. In an A/B experiment switching one of the offline Hadoop based sources of pins for users to the Pixie system improved saves per pin by 50\%.

\subsection{Related Pins}
When a user clicks on a pin at Pinterest, we show Related Pins below.
For example, if a user clicks on a pin containing a trumpet then they can browse
 other pins containing trumpets. Pixie is the primary source of candidates
for Related Pins, though the full system features several additional layers~\cite{relatedpins}. The queries arising in this application
contain a single pin that the user is viewing. This application requires the recommended pins to be very similar to the original query pin. For example, we would prefer that the Related Pins of a trumpet contain other trumpets and not some other musical instrument.
One general observation has been that as the length of random walk increases, Pixie visits pins that are increasingly diverse. Therefore, for Related Pins we hypothesised that users would prefer shorter walks compared to Homefeed.
In fact, we ran online A/B tests on users which showed that by simply decreasing the walk length lead to a significant lift in engagement---the number of Related Pins saved per day increased by 3\%. 

Similarly, we use Pixie to generate pins related to a board.
When users view their own board, Pinterest suggests other pins they can save to this board.
The Pixie query here consist of the last ten pins added to the board. This application helps
the users grow their boards and have better collections.

\subsection{Recommending Boards}
\label{sec:boardrecs}

Another important recommendation problem at Pinterest is ``Picked For
You'' (PFY) boards. In the original Pinterest ecosystem users manually
followed other user's boards to have new pins added to the
follower's stream as they were added to the boards, similar to other
following based systems. This manual approach requires a lot of work
from users to find and maintain the list of boards interesting to them
for following and frequently users would not manually follow enough
boards to get a good feed of new content or would not update them as their
tastes or interest changed leading to reduced engagement with pins.

Here Pixie recommends boards and then delivers the most
recent new pins from those boards to a user's homefeed. This approach has the benefit of providing
additional diversity and a natural distribution of cold-start, new, and trending
content on the site. 

\hide{
Previously a separate offline pipeline was used to generate board
recommendations. It was computationally prohibitive to score users vs. all
possible boards due to the size of users, pins, and boards
so a series of hand-tuned heuristics was employed to generate board
recommendations. Pixie enables automatic board recommendations and
this wal helps with the cold-start problem

When recommending PFY boards there are additional considerations
compared to pin recommendations but we can use the approaches already
described in this paper to customize the random walks for boards. For
example one important consideration when recommending PFY boards is
the rate at which new pins appear on the boards, referred to as {\it
flow} from a board. Following boards which have been abandoned or have
very few new pins added to them is uninteresting for the user. We can
use the same biased walk approach used to recommend pins that prefer
local language matching pins described earlier (Section~\ref{sec:local}) to
prefer boards that have enough flow for following. Other algorithmic
modifications like scaling by degree (Section~\ref{sec:scaledegree})
and multiple pin hit boosting (Section~\ref{sec:multihit}) have also
been shown to improve engagement in our production systems for PFY
board recommendations. 
}

Using Pixie for board recommendations has 
allowed us to deprecate the old offline systems while the A/B
experiment also shows that saves per pin improved by 26\%.

These are only few examples of applications that use Pixie---others include the email and personalized articles. Over half of all the pins that users {\em save} each day on Pinterest come from systems backed by Pixie.

\section{Conclusion}
\label{sec:conclusion}

In this paper we presented Pixie, a flexible graph-based real-time recommender system that we built and deployed at Pinterest. Each server in the Pixie fleet holds the entire bipartite graph of over a billion pins and boards, and supports 1,200 queries per second with a 99-percentile latency of 60 milliseconds. 

We have implemented and deployed the novel Pixie Random Walk algorithm. Our offline experiments have empirically demonstrated high-quality of recommendations, robustness as well as the efficiency of the algorithm. Furthermore, online A/B experiments have been used to launch Pixie to power multiple Pinterest surfaces, most notably the Homefeed and Related Pins products, so that now over half of all pins saved on Pinterest each day come from systems backed by Pixie. We have also found Pixie useful for performing other tasks like label-propagation on the order of minutes instead of days using distributed systems like Hadoop. 

Thanks to the high performance, scalability, and generic nature of the Pixie architecture and algorithms, we anticipate a bright future with new algorithms and graphs featuring novel node types and edge definitions for even more applications at Pinterest.

\xhdr{Acknowledgments}
Allan Blair, Jeremy Carroll, Collins Chung, Yixue Li, David Liu, Jenny Liu, Peter Lofgren, Kevin Ma, and Stephanie Rogers among many who helped make Pixie a success!

\bibliographystyle{abbrv}
{\tiny
\bibliography{refs}

\begin{thebibliography}{10}

\bibitem{linkedin}
D.~Agarwal, B.~Chen, Q.~He, Z.~Hua, G.~Lebanon, Y.~Ma, P.~Shivaswamy, H.~Tseng,
  J.~Yang, and L.~Zhang.
\newblock Personalizing linkedin feed.
\newblock In {\em {KDD}}, pages 1651--1660, 2015.

\bibitem{backstrom2011supervised}
L.~Backstrom and J.~Leskovec.
\newblock Supervised random walks: predicting and recommending links in social
  networks.
\newblock In {\em {WSDM}}, pages 635--644, 2011.

\bibitem{bahmani2010fast}
B.~Bahmani, A.~Chowdhury, and A.~Goel.
\newblock Fast incremental and personalized pagerank.
\newblock {\em PVLDB}, 4(3):173--184, 2010.

\bibitem{baluja2008video}
S.~Baluja, R.~Seth, D.~Sivakumar, Y.~Jing, J.~.Yagnik, S.~Kumar,
  D.~Ravichandran, and M.~Aly.
\newblock Video suggestion and discovery for youtube: taking random walks
  through the view graph.
\newblock In {\em {WWW}}, pages 895--904, 2008.

\bibitem{bennett2007netflix}
J.~Bennett and S.~Lanning.
\newblock The {Netflix} prize.
\newblock In {\em In {KDD} Cup and Workshop in conjunction with {KDD}}, 2007.

\bibitem{cheng2016wide}
H.~Cheng, L.~Koc, J.~Harmsen, T.~Shaked, T.~Chandra, H.~Aradhye, G.~Anderson,
  G.~Corrado, W.~Chai, M.~Ispir, et~al.
\newblock Wide \& deep learning for recommender systems.
\newblock In {\em DLRS Workshop}, pages 7--10, 2016.

\bibitem{youtube}
P.~Covington, J.~Adams, and E.~Sargin.
\newblock Deep neural networks for youtube recommendations.
\newblock In {\em {RecSys}}, pages 191--198, 2016.

\bibitem{covington2016deep}
P.~Covington, J.~Adams, and E.~Sargin.
\newblock Deep neural networks for youtube recommendations.
\newblock In {\em RecSys}, pages 191--198, 2016.

\bibitem{google}
A.~Das, M.~Datar, A.~Garg, and S.~Rajaram.
\newblock Google news personalization: scalable online collaborative filtering.
\newblock In {\em {WWW}}, pages 271--280, 2007.

\bibitem{davidson2010youtube}
J.~Davidson, B.~Liebald, J.~Liu, P.~Nandy, T.~V. Vleet, U.~Gargi, S.~Gupta,
  Y.~He, M.~Lambert, B.~Livingston, and D.~Sampath.
\newblock The youtube video recommendation system.
\newblock In {\em {RecSys}}, pages 293--296, 2010.

\bibitem{van2013deep}
A.~V. den Oord, S.~Dieleman, and B.~Schrauwen.
\newblock Deep content-based music recommendation.
\newblock In {\em NIPS}, pages 2643--2651, 2013.

\bibitem{donahue2014decaf}
J.~Donahue, Y.~Jia, O.~Vinyals, J.~Hoffman, N.~Zhang, E.~Tzeng, and T.~Darrell.
\newblock Decaf: A deep convolutional activation feature for generic visual
  recognition.
\newblock In {\em International conference on machine learning}, pages
  647--655, 2014.

\bibitem{fogaras2005towards}
D.~Fogaras, B.~R{\'a}cz, K.~Csalog{\'a}ny, and T.~Sarl{\'o}s.
\newblock Towards scaling fully personalized pagerank: Algorithms, lower
  bounds, and experiments.
\newblock {\em Internet Mathematics}, 2(3):333--358, 2005.

\bibitem{goel2015follow}
A.~Goel, P.~Gupta, J.~Sirois, D.~Wang, A.~Sharma, and S.~Gurumurthy.
\newblock The who-to-follow system at twitter: Strategy, algorithms, and
  revenue impact.
\newblock {\em Interfaces}, 45(1):98--107, 2015.

\bibitem{wtf}
P.~Gupta, A.~Goel, J.~J. Lin, A.~Sharma, D.~Wang, and R.~Zadeh.
\newblock {WTF:} the who to follow service at twitter.
\newblock In {\em {WWW}}, pages 505--514, 2013.

\bibitem{kabiljo_recommending_2015}
M.~Kabiljo and A.~Ilic.
\newblock Recommending items to more than a billion people.

\bibitem{konstan1997grouplens}
J.~A. Konstan, B.~N. Miller, D.~Maltz, J.~L. Herlocker, L.~R. Gordon, and
  J.~Riedl.
\newblock Grouplens: applying collaborative filtering to usenet news.
\newblock {\em Communications of the ACM}, 40(3):77--87, 1997.

\bibitem{koren2009matrix}
Y.~Koren, R.~M. Bell, and C.~Volinsky.
\newblock Matrix factorization techniques for recommender systems.
\newblock {\em {IEEE} Computer}, 42(8):30--37, 2009.

\bibitem{salsa}
R.~Lempel and S.~Moran.
\newblock {SALSA:} the stochastic approach for link-structure analysis.
\newblock {\em {ACM} Trans. Inf. Syst.}, 19(2):131--160, 2001.

\bibitem{snap}
J.~Leskovec and R.~Sosic.
\newblock {SNAP:} {A} general-purpose network analysis and graph-mining
  library.
\newblock {\em {ACM} {TIST}}, 8(1):1:1--1:20, 2016.

\bibitem{amazon}
G.~Linden, B.~Smith, and J.~York.
\newblock Amazon.com recommendations: Item-to-item collaborative filtering.
\newblock {\em {IEEE} Internet Computing}, 7(1):76--80, 2003.

\bibitem{relatedpins}
D.~Liu, S.~Rogers, R.~Shiau, D.~Kislyuk, K.~Ma, Z.~Zhong, J.~Liu, and Y.~Jing.
\newblock Related pins at pinterest: The evolution of a real-world recommender
  system.
\newblock In {\em {WWW}}, 2017.

\bibitem{mikolov2013distributed}
T.~Mikolov, I.~Sutskever, K.~Chen, G.~Corrado, and J.~Dean.
\newblock Distributed representations of words and phrases and their
  compositionality.
\newblock pages 3111--3119, 2013.

\bibitem{pazzani2007content}
M.~J. Pazzani and D.~Billsus.
\newblock Content-based recommendation systems.
\newblock In {\em The adaptive web}, pages 325--341. Springer, 2007.

\bibitem{sarwar2001item}
B.~Sarwar, G.~Karypis, J.~Konstan, and J.~Riedl.
\newblock Item-based collaborative filtering recommendation algorithms.
\newblock In {\em WWW}, pages 285--295, 2001.

\bibitem{graphjet}
A.~Sharma, J.~Jiang, P.~Bommannavar, B.~Larson, and J.~J. Lin.
\newblock Graphjet: Real-time content recommendations at twitter.
\newblock {\em {PVLDB}}, 9(13):1281--1292, 2016.

\bibitem{simonyan2014very}
K.~Simonyan and A.~Zisserman.
\newblock Very deep convolutional networks for large-scale image recognition.
\newblock {\em arXiv preprint arXiv:1409.1556}, 2014.

\bibitem{Tong:2006:FRW:1193207.1193363}
H.~Tong, C.~Faloutsos, and J.~Pan.
\newblock Fast random walk with restart and its applications.
\newblock In {\em Proceedings of the Sixth International Conference on Data
  Mining}, ICDM '06, pages 613--622, 2006.

\bibitem{zheng2017joint}
L.~Zheng, V.~Noroozi, , and P.~S. Yu.
\newblock Joint deep modeling of users and items using reviews for
  recommendation.
\newblock In {\em WSDM}, pages 425--434, 2017.

\bibitem{zhuang2013fast}
Y.~Zhuang, W.~Chin, Y.~Juan, and C.~Lin.
\newblock A fast parallel sgd for matrix factorization in shared memory
  systems.
\newblock In {\em RecSys}, pages 249--256, 2013.

\end{thebibliography}
}


\end{document}